\def\dst#1{{\displaystyle{#1}}}
\def\r#1{\mbox{\bf r}_{#1}}
\def\rp#1{\mbox{\bf r}_{#1}^{\prime}}
\def\d{\mbox{d}}

\newcommand{\tr}{\mathop{\mathrm{Tr}}\limits}
\documentclass[pre,aps,twocolumn,showpacs,superscriptaddress]{revtex4}
\usepackage{amsmath}
\usepackage{amssymb}
\usepackage{bm}

\begin{document}
\bibliographystyle{apsrev}

\title{Condensate fluctuations in the dilute Bose gas}

\author{Alexander Yu. Cherny}
\email{cherny@mpipks-dresden.mpg.de}
\affiliation{\mbox{Max-Planck-Institut f\"ur Physik komplexer Systeme, N\"othnitzer Stra{\ss}e 38, D-01187 Dresden,
Germany}}
\affiliation{Bogoliubov Laboratory of Theoretical Physics, Joint Institute for Nuclear Research, 141980,
Dubna, Moscow region, Russia}

\date{\today}

\begin{abstract}
The fluctuations of a number of particles in the Bose-Einstein condensate are
studied in the grand canonical ensemble with an effective single-mode
Hamiltonian, which is derived from an assumption that the mode corresponding to
the Bose-Einstein condensate does not asymptotically correlate with other
modes. The fluctuations are evaluated in the dilute limit with a proposed
simple method, which is beyond the mean-field approximation. The accuracy of
the latter is estimated; it is shown that the mean-field scheme does not work
for the single-mode Hamiltonian, while for the Hartree Hamiltonian it allows us
to estimate the condensate fluctuations up to a numerical factor. As a
hypothesis, a formula is proposed that
relates the fluctuations in the canonical ensemble with that of the grand canonical one.
\end{abstract}
\pacs{03.75.Hh, 05.30.Jp}
\maketitle

\section{Introduction}
\label{sec:intro}

The observation of Bose-Einstein condensation (BEC) in trapped alkali-metal
gases~\cite{anddavbrad} has stimulated theoretical and experimental studies of
the basic problems related to this phenomenon (see the reviews
in~\cite{leggett,dalfovo,walls}). In particular, the statistical properties of
the condensate are of especial interest for interacting bosons. In the case of
the ideal Bose gas, fluctuations of a number of particles in the Bose-Einstein
condensate $\langle\delta\hat{N}_{0}^{2}\rangle \equiv
\langle\hat{N}_{0}^{2}\rangle - \langle\hat{N}_{0}\rangle^{2}$ have been
studied thoroughly in a box~\cite{box-id} and a harmonic trap~\cite{trap-id} as
well. However, in the case of the interacting gas, this problem is rather
subtle and requires a more refined approach. So far there has been no uniform
treatment of this problem in the
literature~\cite{jak,dunn,navez,pit,idz,illu,mei,koch,grah,xio,xio1,xio2,bhaduri}.
It is not even clear what the value of the power $\gamma$ is in dependence of
the fluctuations on the total number of particles,
\begin{equation}\label{gamma}
\langle\delta\hat{N}_{0}^{2}\rangle\propto N^{\gamma},
\end{equation}
in various Gibbs ensembles (compare, e.g., the different results of
Refs.~\cite{pit} and \cite{idz,illu}). Apart from  general theoretical
interest, condensate fluctuations can be measured, in principle, by means of a
scattering of series of short laser pulses~\cite{idz1}.

In this paper we deal with the problem of fluctuations in the grand canonical
ensemble and consider the global gauge symmetry $U(1)$ of the Bose system to be
broken. In principle, the properties of the dilute Bose gas in the grand
ensemble are described quite well by the Bogoliubov model~\cite{bog47}, with
the condensate terms being treated as asymptotic $c$-numbers and with three-
and four-boson terms being neglected in the Hamiltonian. However, the
statistical properties of the Bose-Einstein condensate are beyond the
Bogoliubov theory due to the $c$-number replacement. Consequently, in order to
tackle this problem, we need other approximations or assumptions that keep the
quantum nature of the condensate operators. Assuming that the condensate mode
and other modes can  be considered as quasi-independent, we obtain the usual
thermodynamic  fluctuations of the condensate with $\gamma=1$ in
Eq.~(\ref{gamma}) and show that this result is consistent quite well with the
Bogoliubov theory.

We consider a dilute Bose gas interacting with a short-range pairwise potential
[i.e., the potential that goes to zero at $r\to\infty$ as $1/r^m$ ($m>3$) or
faster]. The method used in our paper can be called the approximation of the
quasi-independent mode. This assumption is nothing else but a generalization of
Bogoliubov's relation~(\ref{corrw}) for the operators $\hat{a}_{0}$ and
$\hat{a}_{0}^{\dag}$ (see Sec.~\ref{sec:gen} below). As a result, we consider
the condensate mode and the other ones to be uncorrelated. In the framework of
this approximation, one can easily derive the effective single-mode
Hamiltonian. Our consideration is primarily dedicated to the homogeneous case,
except for Sec.~\ref{sec:mf}, where fluctuations in the Hartree model are
discussed briefly in the nonhomogeneous case.

The paper is organized as follows. In Sec.~\ref{sec:gen} we review some
important issues in the field of Bose-Einstein condensation, which might not be
well known  for a reader. In the next section we derive the effective 
single-mode Hamiltonian, which is employed for studying condensate fluctuations
in the grand canonical  ensemble. In Secs.~\ref{sec:diff} and \ref{sec:mf} the
fluctuations are  evaluated in the framework of the mean-field approximation
and beyond it. Possible corrections to the value of  the fluctuations are
discussed in Sec.~\ref{sec:corr}. In the last  section the main results are
summarized. 

\section{General remarks}
\label{sec:gen}

Let us recall some important issues in the considered problem, following
Bogoliubov's paper~\cite{bogquasi}. A number of bosons in the Bose-Einstein
condensate can be defined as the macroscopic eigenvalue $N_0$ of the one-body
density matrix $\langle \hat{\psi}^{ \dag} ({\bf r}) \hat{\psi}({\bf
r}')\rangle$ -- that is, $\int \d^3r' \, \langle \hat{\psi}^{ \dag} ({\bf r}')
\hat{\psi}({\bf r})\rangle \phi_{0}({\bf r}') = N_{0} \phi_{0}({\bf r})$, $\int
\d^3r\,|\phi_{0}({\bf r})|^{2}=1$, where $N_0$ is proportional to the total
number of bosons in the thermodynamic limit, while the other eigenvalues are
proportional to the unit. The eigenfunction $\phi_{0}$ is a one-body wave
function of the Bose-Einstein condensate, obeying the Gross-Pitaevskii equation
in the first approximation. One can easily introduce the condensate operators
$\hat{a}_{0}$ and $\hat{a}_{0}^{\dag}$ by expanding the Bose field operators in
the complete set of eigenfunctions of the one-body matrix: $\hat\psi({\bf
r})=\hat{a}_{0}\phi_0({\bf r})+\sum_{j\not=0}\hat{a}_{j}\phi_{j}({\bf r})$;
then, the operator $\hat{N}_0=\hat{a}_{0}^{\dag}\hat{a}_{0}$ describes the
number of particles in the condensate: $\langle\hat{N}_0\rangle=N_0$. For a
homogeneous system the index $j$ is associated with momentum ${\bf p}$, 
$\phi_{\bf p}({\bf r})=\exp(i{\bf p}\cdot{\bf r})/\sqrt{V}$ (V is the volume),
and the condensate operators coincide with the Bose zero-momentum operators.
Bogoliubov noticed~\cite{bog47,bogquasi} that in this case the operators
$\hat{a}_{0}/\sqrt{V}$ and $\hat{a}_{0}^{\dag}/\sqrt{V}$ should be very close
to $c$-numbers, since they commute in the thermodynamic limit $V\to\infty$,
$n=N/V={\rm const}$. This implies that
\begin{equation}\label{asymp}
\lim_{V\to\infty}\left\langle\bigg(\frac{\hat{a}_{0}^{\dag}}{\sqrt{V}}-\sqrt{n_0}e^{-i\varphi}\bigg)
\bigg(\frac{\hat{a}_{0}}{\sqrt{V}}-\sqrt{n_0}e^{i\varphi}\bigg)\right\rangle=0,
\end{equation}
where $\varphi$ is an arbitrary phase and $n_0=N_0/V$ denotes the density of
the Bose-Einstein condensate. However, we always have the constraint
$\langle\hat{a}_{0}\rangle=\langle\hat{a}_{0}^{\dag}\rangle=0$ due to global
gauge invariance, which is equivalent to conservation of the total number of
bosons (i.e., $[\hat{H},\hat{N}]=0$). Hence, the ground state is not stable
with respect to an infinitesimally small perturbation that breaks the gauge
symmetry. For a correct mathematical treatment of the homogeneous Bose system
with broken symmetry, Bogoliubov proposed~\cite{bogquasi} to include the terms
$-\nu(\hat{a}_{0}^{\dag}e^{i\varphi} + \hat{a}_{0}e^{-i\varphi})\sqrt{V}$ in
the Hamiltonian, where the parameter $\nu>0$ and $\nu\to0$. Now the absolute
minimum of the energy corresponds to the state with
$\langle\hat{a}_{0}\rangle/\sqrt{V}=
\langle\hat{a}_{0}^{\dag}\rangle^{*}/\sqrt{V} = \sqrt{n_0}e^{i\varphi}$,
because the gain in energy per particle, by Eq.~(\ref{asymp}), is equal to
$2\nu\sqrt{n_0/n}$ in the limit $V\to\infty$ due to Bogoliubov's terms. We
stress that the limit $\nu\to0$ should be performed {\it after} the
thermodynamic one. These two subsequent limits yield the well-defined order
parameter $\langle\hat{\psi}({\bf r})\rangle =
\langle\hat{a}_{0}\rangle/\sqrt{V}=\sqrt{n_0}e^{i\varphi}$ with fixed phase
$\varphi$. Thus, due to Bogoliubov's  infinitesimal terms, the values of the
{\it anomalous} averages (like $\langle\hat{\psi}\rangle$,
$\langle\hat{\psi}\hat{\psi}\rangle$, and so on) change drastically, and such
averages  can be called, following Bogoliubov, quasiaverages (hereafter by the
term ``average'' we mean ``quasiaverage'').  At the same time, the values of
the {\it normal} averages  (like $\langle\hat{\psi}^{\dag}\hat{\psi}\rangle$
and so on) do not change. In particular, this is valid for the one-body
$\langle \hat{\psi}^{ \dag} ({\bf r}) \hat{\psi}({\bf r}')\rangle$ and two-body
$\langle{\hat\psi}^{\dag}(\r{1}) {\hat\psi}^{\dag}(\r{2}) {\hat\psi}(\rp{2})
{\hat\psi}(\rp{1})\rangle$ matrices. Hence, in the thermodynamic limit the
eigenfunctions of the matrices do not change when the symmetry is broken. On
the other hand, the anomalous averages have a transparent physical
interpretation: $\langle\hat\psi({\bf r})\rangle=\sqrt{N_{0}}\phi_{0}({\bf r})$
is the eigenfunction of the one-body matrix associated with the maximum
eigenvalue $N_0$, and $\langle\hat\psi({\bf r})\hat\psi({\bf r}')\rangle$ is
that of the two-body matrix with the eigenvalue $N_0(N_0-1)$~\cite{cherny}.
Hence, with the concept of broken symmetry we obtain a simple method of
evaluating these eigenfunctions.

Equation~(\ref{asymp}) can be derived from Bogoliubov's principle of
correlation weakening~\cite{bogquasi}, which takes a particular form
$\langle\hat{\psi}({\bf r})\cdots \hat{\psi}^{\dag}({\bf r}_1)\cdots
{\hat\psi}({\bf r}_2) \cdots\rangle\simeq\langle\hat{\psi}({\bf r})\rangle
\langle\cdots \hat{\psi}^{\dag}({\bf r}_1)\cdots  {\hat\psi}({\bf r}_2)
\cdots\rangle$ when $|{\bf r}-{\bf r}_{1}| \to \infty$, $\cdots$ and  $|{\bf
r}-{\bf r}_{2}| \to \infty$. Indeed, using the expression of the condensate
operator in the homogeneous system,  $\hat{a}_{0}=\int\d^{3}r\,\hat{\psi}({\bf
r})/\sqrt{V}$, we obtain, in the thermodynamic limit,
\begin{align}
\bigg\langle\frac{\hat{a}_{0}}{\sqrt{V}}&\cdots \hat{\psi}^{\dag}({\bf r}_1)\cdots {\hat\psi}({\bf r}_2)
\cdots\bigg\rangle \nonumber\\
=&\frac{1}{V}\int\d^{3}r\,\left\langle\hat{\psi}({\bf r})\cdots \hat{\psi}^{\dag}({\bf
r}_1)\cdots {\hat\psi}({\bf r}_2) \cdots\right\rangle \nonumber\\
=&\frac{1}{V}\int\d^{3}r\,\left\langle\hat{\psi}({\bf r})\right\rangle \left\langle\cdots
\hat{\psi}^{\dag}({\bf r}_1)\cdots  {\hat\psi}({\bf r}_2) \cdots\right\rangle \nonumber\\
=&\frac{\langle\hat{a}_{0}\rangle}{\sqrt{V}}
\left\langle\cdots{\hat\psi}^{\dagger}({\bf r}_1)\cdots{\hat\psi}({\bf r}_2)
\cdots\right\rangle.
\label{corrw}
\end{align}
One can indeed replace the condensate operators by the $c$-numbers, and this
procedure, according to Bogoliubov, still yields {\it exact}
values~\cite{bogquasi} of all thermodynamic quantities in the limit
$V\to\infty$. Note that the above relations can easily be extended to the
inhomogeneous case. For example, in order to consider the harmonic trap, it is
sufficient to replace the volume $V$ by $1/\bar{\omega}^{3}$ and define the
thermodynamic limit as~\cite{damle} $N\to\infty$, $N\bar{\omega}^{3}={\rm
const}$ (here $\bar{\omega}=\sqrt[3]{\omega_{x} \omega_{y} \omega_{z}}$ is the
average harmonic frequency). 

We stress once more that the replacement of $\hat{a}_{0}$ and
$\hat{a}^\dag_{0}$ by the $c$-numbers gives always the main asymptotics of the
averages of these operators (if the Bose-Einstein condensate exists), and this
method is sufficient for obtaining an exact value of any thermodynamic quantity
(see Sec.~7 of Bogoliubov's paper~\cite{bogquasi}). However, the main
asymptotic value is canceled when evaluating the condensate fluctuations
$\langle\delta\hat{N}_{0}^{2}\rangle$, and, thus, we have to take into
consideration here the quantum nature of the condensate operators.
Nevertheless, Eq.~(\ref{asymp}) results in the constraint $\gamma<2$ for the
parameter $\gamma$ in Eq.~(\ref{gamma}). The case $\gamma>1$ implies that
bosons in the Bose-Einstein condensate and beyond it are strongly correlated,
and the extent of the correlation is nonthermodynamically large, since the
thermodynamic fluctuations of any extensive observable are proportional to the
total number of particles. The results $\gamma=4/3$ and $\gamma=1$ have been
obtained in Refs.~\cite{pit} and \cite{idz,illu}, respectively, in the Gibbs
canonical ensemble. The results of Refs.~\cite{idz,illu} are consistent with
ours for the grand canonical ensemble, while those of Ref.~\cite{pit} are at
variance with ours (see Sec.~\ref{sec:corr} below).

If the symmetry is broken (that is, if the ground state is a superposition of
states with different numbers of particles), then it is rather difficult to
define what the Gibbs canonical ensemble is. Indeed, within a variational
scheme, we cannot use the restriction $N={\rm const}$ directly and have to
impose the additional condition $\langle\hat{N}\rangle=N$, which is equivalent
to introducing the Lagrange term $-\mu\hat{N}$ in the Hamiltonian; this is
nothing else but using the Gibbs grand canonical ensemble. So, a reliable
treatment of the canonical ensemble can be done in the framework of the
number-conserving scheme (for recent developments of the scheme, see
Refs.~\cite{gard,cast}). For that reason, in this paper we restrict ourselves
to the grand ensemble.

\section{Approximation of the quasi-independent mode}
\label{sec:model}

Let us describe the model and methods of calculation. In order to evaluate the
mean square of the number fluctuations in the Bose-Einstein condensate, it
sufficient to know the condensate density matrix $\hat{\rho}_{\rm c}$.
According to  general rules of quantum mechanics, it can be obtained from the
total density matrix $\hat{\rho}$ by taking the partial trace in the occupation
number representation
\begin{equation}
\hat{\rho}_{\rm c}=\tr_{\cdots n_p\cdots\atop p\not=0}\hat{\rho}=\sum_{\cdots n_p\cdots\atop p\not=0} \langle\cdots
n_{p}\cdots|\hat{\rho}|\cdots n_{p}\cdots\rangle,
\label{rhoc}
\end{equation}
where we denote $|\cdots n_{p}\cdots\rangle=\prod_{p\not=0}|n_{p}\rangle$ and
the index $p$ is associated with momentum in the homogeneous case. So
$\hat{\rho}_{\rm c}$ depends on the condensate operators $\hat{a}_{0}^{\dag}$
and $\hat{a}_{0}$ only and acts only on  states of the condensate -- say, the
Fock states $|N_0\rangle$ and their superpositions. By analogy with
Eq.~(\ref{rhoc}), one can also define the density matrix for all noncondensate
states $\hat{\rho}_{\rm out}=\tr_{N_0}\hat{\rho}$, which depends on the Bose
operators $\hat{a}_{p}^{\dag}$ and $\hat{a}_{p}$ with $p\not=0$. Certainly, it
is difficult to evaluate the density matrix of the Bose-Einstein condensate
directly from Eq.~(\ref{rhoc}), and one has to use additional assumptions.

As discussed in Sec.~\ref{sec:gen}, Eq.~(\ref{corrw}), a particular case of
Bogoliubov's principle of correlation weakening, is valid in the thermodynamic
limit. This suggests that the state of the condensate mode is, in effect,
independent of the other modes. It means that one can use the approximation for
the density matrix $\hat{\rho}\simeq \hat{\rho}_{\rm c}\hat{\rho}_{\rm out}$.
The parameters of the matrices $\hat{\rho}_{\rm c}$ and $\hat{\rho}_{\rm out}$
are assumed to be related in a self-consistent manner. The factorization of the
total density matrix implies that the condensate and noncondensate  subsystems
are considered to be uncorrelated. Hence, we have
$\langle\hat{A}\hat{B}\rangle\simeq\langle\hat{A}\rangle\langle\hat{B}\rangle$
for arbitrary $\hat{A}$ and and $\hat{B}$, depending on the condensate and
noncondensate operators, respectively. This is very consistent with the
approximation~\cite{solomon}
\begin{equation}
\hat{A}\hat{B} \simeq \langle\hat{A}\rangle \hat{B} + \hat{A} \langle\hat{B}\rangle
- \langle\hat{A}\rangle\langle\hat{B}\rangle.
\label{decoupl}
\end{equation}

Let us apply the approximation~(\ref{decoupl}) for a homogeneous Bose system
with the pairwise potential $V(r)$. In this case, the Hamiltonian reads, in
terms of the Bose fields operators,
\begin{align}
\hat H\!=&\int\d^{3}r\,{\hat\psi}^{\dagger}({\bf r})
\bigg(\!-\frac{\hslash^2\nabla^{2}}{2m}-\mu\bigg){\hat\psi}({\bf r})\nonumber\\
&+\int\d^{3}r\d^{3}r'\,V(|{\bf r}-{\bf r}'|)
{\hat\psi}^{\dagger}({\bf r}){\hat\psi}^{\dagger}({\bf r}')
{\hat\psi}({\bf r}'){\hat\psi}({\bf r}).
\label{ham}
\end{align}
According to the above method, one should separate explicitly the condensate
operators in the Hamiltonian by means of substitution ${\hat\psi}({\bf r})
=\hat{a}_{0}/\sqrt{V} +{\hat\vartheta}({\bf r})$ [here we denote
${\hat\vartheta}({\bf r})=\sum_{p\not=0}\hat{a}_{\bf p}\exp(i{\bf p}\cdot{\bf
r})/\sqrt{V}$] and  employ the decomposition (\ref{decoupl}) to every  operator
containing the product of the condensate operators and  $\hat{\vartheta}$ or
$\hat{\vartheta}^{\dag}$. As one can see, applying Eq.~(\ref{decoupl}) to the
Hamiltonian~(\ref{ham}), we come to the approximation
\begin{equation}
\hat{H} \simeq\hat{H}_{\rm c} +\hat{H}_{\rm out}+\mbox{\rm const},
\label{ham1}
\end{equation}
and, hence, the density matrix of the grand canonical ensemble 
$\hat{\rho}\sim\exp(-\hat{H}/T)$ is really factorized (in this paper we include
the term $-\mu\hat{N}$ in the Hamiltonian). By virtue of this factorization,
the condensate fluctuations arise in the grand canonical ensemble due to the
particle and temperature bath rather than the energy excitations. Here we
derive the effective grand canonical Hamiltonian of the Bose-Einstein
condensate:
\begin{eqnarray}
\hat{H}_{\rm c}&=&\frac{1}{2}\Big(V_{0}\frac{\hat{a}_{0}^{\dag2}{\hat{a}_{0}}^2}{V}
+A(\hat{a}_{0}^{\dag2} + {\hat{a}_{0}}^2) +2(B-\mu)\hat{a}_{0}^{\dag}\hat{a}_{0} \nonumber\\
&&+2C(\hat{a}_{0}^{\dag} + \hat{a}_{0})\sqrt{V}\Big),
\label{heff}
\end{eqnarray}
where $V_0$ is the zero component of the Fourier transform of the pairwise
interaction $V(r)$, and we introduce by definition the coefficients
\begin{eqnarray}
A&=&\int \d^{3}r\,V(r)\langle\hat{\vartheta}({\bf r})\hat{\vartheta}(0)\rangle, \nonumber\\
B&=&V_0\langle\hat{\vartheta}^{\dag}(0)\hat{\vartheta}(0)\rangle
   +\int \d^{3}r\, V(r)\langle\hat{\vartheta}^{\dag}({\bf r})\hat{\vartheta}(0)\rangle, \nonumber\\
C&=&\int \d^{3}r\,V(r)\langle\hat{\vartheta}^{\dag}({\bf r})\hat{\vartheta}({\bf r})\hat{\vartheta}(0)\rangle.
\label{coeff}
\end{eqnarray}
In the above expressions, the gauge symmetry is certainly assumed to be broken,
and for the  sake of simplicity, we put $\varphi=0$ in Bogoliubov's terms (see
Sec.~\ref{sec:gen}); in this case, all the coefficients are real. In general,
they depend on the density and temperature and will be evaluated below. It is
worthwhile to note that $A$, $B$, and $C$ tend to constant values in the
thermodynamic limit. As to the Hamiltonian $\hat{H}_{\rm out}$, it is given by
Eq.~(\ref{ham}) but with $c$-numbers instead of the condensate
operators~\cite{note6}, so we arrive at the standard asymptotically exact
Hamiltonian with Bogoliubov's replacement. The coefficients (\ref{coeff}) can
be calculated by means of the Hamiltonian $\hat{H}_{\rm out}$.
 
When calculating the coefficients in the dilute limit, one can employ the
results of our previous papers~\cite{cherny,cherny1,cherny2,cherny3}, based on
an analysis of the two-body density matrix. One can show the following.\\
(i) The functions 
\begin{equation}
\varphi(r)=1+\psi(r),\;
\varphi_{{\bf p}}({\bf r})= \sqrt{2}\cos({\bf p}\cdot{\bf r})+
\psi_{{\bf p}}({\bf r})\quad (p\not=0)
\label{phi}
\end{equation}
are eigenfunctions of the two-body density matrix, which belong to the
continuous spectrum and describe the scattering of pairs of bosons in a medium
of other bosons. The quantum number $p$ can be associated with the relative
momentum of that scattering [$\varphi(r)$ corresponds to zero momentum]. Their
scattering parts $\psi(r)$ and $\psi_{{\bf p}}$, respectively, obey  the
boundary conditions $\psi(r),\;\psi_{{\bf p}} ({\bf r}) \to 0$ at $r \to
\infty$.  The Fourier transforms of the scattering parts can be expressed in
terms of the Bose operators:
\begin{eqnarray}
\psi(k)=
\frac{\langle \hat{a}_{{\bf k}}\hat{a}_{-{\bf k}}\rangle}{n_0},\
{\psi}_{\bf p}({\bf k})= \sqrt{\frac{V}{2 n_0}}
\frac{\langle \hat{a}^{\dagger}_{2{\bf p}}
\hat{a}_{{\bf p}+{\bf k}} \hat{a}_{{\bf p}-{\bf k}}\rangle}{n_{2p}},
\label{psik}
\end{eqnarray}
where  $n_{p}=\langle \hat{a}_{\bf p}^{\dagger} \hat{a}_{\bf p} \rangle$ stands for the average
occupation number of bosons.\\
(ii) The following limiting relations are valid:
\begin{equation}
\lim_{p \to 0} \varphi_{{\bf p}} ({\bf r})=\sqrt{2}\varphi(r), \quad
\lim_{n \to 0} \varphi(r)=\varphi^{(0)}(r),
\label{limrel}
\end{equation}
where $\varphi^{(0)}(r)$ obeys the two-body Schr\"odinger equation, describing
the scattering of two bosons in a vacuum with zero relative momentum.  Now,
using relations (\ref{psik}) and (\ref{limrel}), one can obtain the
coefficients (\ref{coeff}) in the dilute limit $n\to0$ after a small amount of
algebra~\cite{note3}
\begin{eqnarray}
A&\simeq& n_0[U^{(0)}(0)-V_0],\quad B\simeq2(n-n_0)V_0, \nonumber\\
C&\simeq&2\sqrt{n_0}(n-n_0)[U^{(0)}(0)-V_0].
\label{coeff1}
\end{eqnarray}
Here $U^{(0)}(0)=\int\d^3r\,\varphi^{(0)}(r)V(r)=4\pi\hslash^2a/m$ stands for
the scattering amplitude and $a$ is the scattering length.

Now all averages of the operators $\hat{a}_{0}^{\dag}$ and $\hat{a}_{0}$ and
their products can be calculated in the grand ensemble by means of the matrix
\begin{equation}
\hat{\rho}_{\rm c}=\exp(-\hat{H}_{\rm c}/T)/Z_c,
\label{rhoc1}
\end{equation}
with the grand partition function for the condensate
\begin{equation}
Z_c=\tr_{N_{0}}\exp(-\hat{H}_{\rm c}/T).
\label{zc}
\end{equation}

Note that the approximation of the quasi-independent mode was introduced
earlier in Ref.~\cite{glass}. The authors considered the quadratic terms only
in operators $\hat{a}$ and $\hat{a}^{\dag}$ in Eq.~(\ref{heff}) (see the
definition in Sec.~\ref{sec:mf} below) and obtained the condensate ground
state, which turns out to be the coherent squeezed state in this case. However,
this approach leads to additional approximations, which change drastically the
final expression for the condensate fluctuations (see below). In order to avoid
the additional approximations,  we need to keep the four-boson Hartree term in
the condensate Hamiltonian (\ref{heff}).

The model~(\ref{heff}) yields the asymptotically exact value of the condensate
density provided that the coefficients~(\ref{coeff}) are known. Indeed, it
follows from Eq.~(\ref{corrw}) that the condensate operators can be replaced by
their mean values $\langle\hat{a}_{0}/\sqrt{V}\rangle
=\langle\hat{a}_{0}^{\dag}/\sqrt{V}\rangle=\sqrt{n_0}$ in order to evaluate the
main asymptotics of $\langle\hat{H}_{\rm c}\rangle$ in the thermodynamic limit.
Then the condensate density can be found by minimization of
$\langle\hat{H}_{\rm c}\rangle$ with respect to the $c$-number parameter $n_0$.
The condition of this minimum can also be obtained from  the initial
Hamiltonian (\ref{ham}), since Bogoliubov's   replacement is asymptotically
exact [22] due to Eq.~(\ref{corrw}). Hence, the condensate density $n_0$ can be
considered as a variational parameter whose value is obtained by minimization
of the thermodynamic potential of the grand canonical ensemble. With the
well-known expression for an infinitesimal change of the potential,
$\delta\Omega= \langle\delta\hat H\rangle$, we have $\partial\Omega/\partial
n_0= \langle\partial\hat H/\partial n_0\rangle=\partial\langle\hat{H}_{\rm
c}\rangle/\partial n_0=0$. Here the second equality is due to the condensate
operators being involved explicitly only in $\hat{H}_{\rm c}$. The last
equation reads
\begin{equation}\label{mu}
\mu=V_0 n_0 + A + B + C/\sqrt{n_0}.
\end{equation}
We come to the equation obtained by Bogoliubov and treated by him as
asymptotically exact [see Eq.~(7.16) of Ref.~\cite{bogquasi}], since its
derivation is based only on the $c$-number replacement of the condensate
operators. It can be rewritten as $\mu=\int\d^3r'\,V(|{\bf r}-{\bf r}'|)
\langle{\hat\psi}^{\dagger}({\bf r}'){\hat\psi}({\bf r}') {\hat\psi}({\bf
r})\rangle/\sqrt{n_0}$, where ${\hat\psi}({\bf
r})=\sqrt{n_0}+{\hat\vartheta}({\bf r})$. This equation relates the equilibrium
value of the condensate density $n_0$ with that of the chemical potential.

\section{Condensate fluctuations}
\label{sec:fluct}

\subsection{Method of parameter differentiation}
\label{sec:diff}

We adopt a standard method of evaluating the fluctuations from the Hamiltonian
(\ref{heff}). One can derive a useful relation by differentiation of the
expression for the average $\langle\hat{N}_{0}\rangle={\rm
Tr}\,(\hat{\rho}_{\rm c}\hat{N}_{0})$ with respect to the chemical potential.
If the Hamiltonian $\hat{H}_{\rm c}$ had commuted with $\hat{N}_{0}$, we would
have $\langle\delta \hat{N}_{0}^{2}\rangle/N_{0} =T({\partial
n_0}/{\partial\mu})_T/{n_0}$ from Eqs.~(\ref{rhoc1}) and (\ref{zc}), but the
noncommutativity leads to corrections we need to estimate. One can apply the
identity $\delta\exp \hat{\Gamma}=\int_0^1\d
z\,\exp[z\hat{\Gamma}]\,\delta\hat{\Gamma}\exp[(1-z) \hat{\Gamma}]$ (here
$\delta$ denotes variation) to the operator $\hat{\Gamma} =-\hat{H}_{\rm c}/T$
and use the first two terms of the expansion
\begin{equation}\label{en0e}
e^{z\hat{\Gamma}}\hat{N}_{0}e^{-z\hat{\Gamma}}=\hat{N}_{0}+[\hat{\Gamma},\hat{N}_{0}]\frac{z}{1!}+
[\hat{\Gamma},[\hat{\Gamma},\hat{N}_{0}]]\frac{z^2}{2!}+\cdots.
\end{equation}
This yields
\begin{eqnarray}
T\frac{\partial Z}{\partial\mu}&=&T\frac{\partial{\rm Tr}\,e^{\hat{\Gamma}}}{\partial\mu}
                                ={\rm Tr}\,(e^{\hat{\Gamma}}\hat{N}_{0}),\nonumber\\
T\frac{\partial{\rm Tr}\,(e^{\hat{\Gamma}}\hat{N}_{0})}{\partial\mu}&=&\int_0^1\d z\,{\rm Tr}\,\Big
(\Big[\hat{N}_{0}+\frac{A}{T}(\hat{a}_{0}^{\dag2}-\hat{a}_{0}^{2})\nonumber\\
&&+\frac{C}{T}\sqrt{V}(\hat{a}_{0}^{\dag}-\hat{a}_{0})+\cdots\Big]
e^{\hat{\Gamma}}\hat{N}_{0}\Big). \nonumber
\end{eqnarray}
In order to estimate the second term on the r.h.s. of the last equation, we
note that ${\rm Tr}\,(\hat{a}_{0}^{\dag2}e^{\hat{\Gamma}}\hat{N}_{0}) =[{\rm
Tr}\,(\hat{a}_{0}^{\dag2}e^{\hat{\Gamma}}\hat{N}_{0})]^{*} ={\rm
Tr}\,(\hat{N}_{0}e^{\hat{\Gamma}}\hat{a}_{0}^{2})$ owing to the reality of the
coefficients in the Hamiltonian~(\ref{heff}). Here we derive
\[
{\rm Tr}\,\Big((\hat{a}_{0}^{\dag2}-\hat{a}_{0}^{2})e^{\hat{\Gamma}}\hat{N}_{0}\Big)
={\rm Tr}\,\Big(e^{\hat{\Gamma}}[\hat{a}_{0}^{2},\hat{N}_{0}]\Big)=2\langle\hat{a}_{0}^{2}\rangle Z
\]
and, in the same manner,
\[
{\rm Tr}\,\Big((\hat{a}_{0}^{\dag}-\hat{a}_{0})e^{\hat{\Gamma}}\hat{N}_{0}\Big)
=\langle\hat{a}_{0}\rangle Z.
\]
Thus, we arrive at the asymptotic expression
\begin{equation}\label{disp}
\frac{\langle\delta \hat{N}_{0}^{2}\rangle}{N_{0}}=\frac{T}{n_0}\left(\frac{\partial n_0}{\partial\mu}\right)_T
-\frac{2A}{T}-\frac{C}{T}\frac{1}{\sqrt{n_0}}+\cdots.
\end{equation}
We stress that the chemical potential~(\ref{mu}) is involved in implicit form
in the coefficients $A$, $B$, and $C$, but it is clear from the above
consideration that one should hold the coefficients constant when taking the
derivation $\partial\mu/\partial n_0$ in Eq.~(\ref{disp}). As a result, we
obtain, from Eq.~(\ref{mu}),
\begin{equation}\label{dmudn0}
{n_0}(\partial \mu/\partial n_0)_T=V_0 n_0 - C/(2\sqrt{n_0}).
\end{equation}

In the thermodynamic limit, Eq.~(\ref{mu}) is the exact
relation~\cite{bogquasi} for the chemical potential, and it can be
applied~\cite{cherny} at any densities and for arbitrary strong pairwise
potential~\cite{classif}. By contrast, Eq.~(\ref{disp}) cannot be employed in
the strong-coupling regime, for which $V_0\to+\infty$, since the divergence
appearing in the coefficients~(\ref{coeff1}) is not canceled in this
expression~(\ref{disp}) \cite{note}. However, one can use Eqs.~(\ref{disp}) and
(\ref{dmudn0}) in the weak-coupling case, for Eq.~(\ref{en0e}) is in fact the
expansion in terms of the coupling constant. Nevertheless, simple physical
arguments (see Sec.~V in the first paper of Ref.~\cite{cherny1}) can help us to
extend our results to the strong-coupling regime. Since the properties of
dilute quantum gases are ruled by the scattering length, the final expression
for condensate fluctuations should depend on the pairwise potential $V(r)$
through mediation of the scattering length in the strong-coupling case. From
this expression one can derive the formula for the weak-coupling regime by
means of the Born series for the scattering amplitude (length): $U^{(0)}(0)
=4\pi\hslash^2a/m =U_0 + U_1 +\cdots$, where  $U_0=V_0$ and $U_1 =-(2\pi)^{-3}
\int\d^{3}k\,V^{2}_k/(2T_{k})<0$, $T_k=\hslash^2k^2/(2m)$, and $V_k$ is the
Fourier transform of the pairwise interaction. Otherwise, the relation obtained
in the weak-coupling case should involve some first terms of the Born series,
but coefficients before the term $U_0=V_0$ are the same as before $U^{(0)}(0)$
in the strong-coupling case. Thus, in the ``weak-coupling" formulas the
substitution $V_0\to U^{(0)}(0)$ (and $U_1,U_2,\ldots\to0$) should be made to
obtain the ``strong-coupling" formulas. Performing this substitution in the
coefficients~(\ref{coeff1}) yields $A\to0$, $C\to0$, and Eqs.~(\ref{disp}) and
(\ref{dmudn0}) result in a simple final answer:
\begin{equation}\label{flfinal}
\frac{\langle\delta \hat{N}_{0}^{2}\rangle}{N_{0}}=\frac{m}{4\pi\hslash^{2}n_0}\frac{T}{a}.
\end{equation}
This equation is valid for sufficiently small depletion of the condensate --
that is, when $T\ll T_{\text{c}}$. Note that the above consideration allows us
to avoid~\cite{cherny1,cherny2} the divergence $U_1\to-\infty$ arising in the
standard pseudopotential approximation $V_k=4\pi\hslash^2a/m$.

The result~(\ref{flfinal}) is a direct consequence of the single-mode
approximation~(\ref{heff}). Indeed, when deriving the Hamiltonian~(\ref{heff}),
we neglect the correlations between the condensate and noncondensate particles;
besides, the condensate depletion is small for the dilute Bose gas.
Consequently, we can put
\begin{equation}
\langle\delta\hat{N}_{0}^{2}\rangle\simeq \langle\delta\hat{N}^{2}\rangle
=T\left(\frac{\partial N}{\partial \mu}\right)_T\simeq \frac{m V}{4\pi\hslash^{2}}\frac{T}{a}.
\label{fltotal}
\end{equation}
Here the familiar expression is employed for the fluctuations of the total
number of particles in the grand ensemble and the formula for the chemical
potential, $\mu=4\pi\hslash^{2}an/m$, in leading order at the density, which
follows from Eqs.~(\ref{coeff1}) and (\ref{mu}). Note that this expression is
nothing else but the relationship between the compressibility 
$\chi_{T}=(\partial n/\partial P)_{T}/n$ and  the fluctuations of the total
number of particles, where $P$ stands for the pressure. Indeed, with the help
of the  thermodynamic relation $(\partial n/\partial P)_{T}=n(\partial
n/\partial \mu)_{T}$ and the definition of the compressibility, it can be
written in the form
\[
\frac{\langle\delta\hat{N}_{0}^{2}\rangle}{N_{0}}\simeq \frac{\langle\delta\hat{N}^{2}\rangle}{N}
=Tn\chi_{T}.
\]
For the ideal Bose gas $a=0$, and we come to an infinitely large value of the
fluctuations because of the nonphysical behaviour of the compressibility of the
ideal Bose gas~\cite{box-id,trap-id}. Note that the fluctuations of the total
number of particles in the grand canonical ensemble remain finite and approach
zero as $T\to0$ when $a$ is finite, however weak the interaction may be; this
is also valid for the one- and two-dimensional Bose gases~\cite{bhaduri}.

\subsection{Mean-field calculations}
\label{sec:mf}

It is interesting to compare the result~(\ref{flfinal}) with the mean-field
calculations for the condensate fluctuations. By separating the $c$-number part
$\sqrt{z}$ in the condensate operator $\hat{a}_0=\sqrt{z}+\hat{a}$ (hence,
$\langle\hat{N}_{0}\rangle=N_{0}=\langle {\hat a}^{\dag}\hat{a}\rangle+z$), the
fluctuation of the condensate in the grand ensemble can be represented in the
form
\begin{equation}\label{flmf}
\langle\delta\hat{N}_{0}^{2}\rangle=(x+1/2)^{2} + y^{2} -1/4 +2z(x+1/2+y),
\end{equation}
where the notations $x=\langle {\hat a}^{\dag}\hat{a}\rangle$ and $y=\langle
\hat{a}^{\dag}\hat{a}^{\dag}\rangle =\langle \hat{a}\hat{a}\rangle$ are
introduced, and the decoupling $\langle {\hat
a}^{\dag2}\hat{a}^{2}\rangle=2x^2+y^2$ is employed in accordance with Wick's
theorem.

Let us study the Hartree grand canonical Hamiltonian
\begin{equation}
\hat{H}_{\rm h}=\frac{V_{0}}{2V}\hat{a}_{0}^{\dag2}{\hat{a}_{0}}^2-\mu\hat{a}_{0}^{\dag}{\hat{a}_{0}},
\label{hartree}
\end{equation}
which is the model Hamiltonian (\ref{heff}) in the case of $A=B=C=0$. To
evaluate the parameters $x$ and $y$ in Eq.~(\ref{flmf}), which correspond to
the Hamiltonian (\ref{hartree}), one can use the Gibbs-Bogoliubov inequality
\begin{equation}\label{fb}
\Omega\leqslant \Omega_0 +\langle\hat{H}-\hat{H}_0\rangle_0.
\end{equation}
Here $\langle\cdots\rangle_0$ means the averages in the grand Gibbs ensemble
with the Hamiltonian $\hat{H}_0$, and $\Omega$ and $\Omega_0$ are the grand
thermodynamic potentials corresponding to the arbitrary Hamiltonians $\hat{H}$
and $\hat{H}_0$, respectively. Now, the basic idea is to choose
$\hat{H}=\hat{H}_{\rm h}$ and
\begin{equation}\label{h0}
\hat{H}_0=\frac{1}{2}\Big(A_0\hat{a}^{\dag2} + A_0^*\hat{a}^2
+2B_0\hat{a}^{\dag}\hat{a}\Big),
\end{equation}
with arbitrary parameters $A_0$ and $B_0$, and minimize the r.h.s. of
Eq.~(\ref{fb}) with respect to them. As a result, we find the stationary values
of $A_0$ and $B_0$. Note that $B_0$ is always real, and we can put $A_0=A_0^*$
if all coefficients are real in the Hamiltonian $\hat{H}$. By using
Bogoliubov's transformation, one can find the values $x$ and $y$ with the
Hamiltonian~(\ref{h0})
\begin{eqnarray}
x&=&\langle {\hat a}^{\dag}\hat{a}\rangle_0=
\frac{B_0}{2\varepsilon}\coth\frac{\varepsilon}{2T}-\frac{1}{2},\nonumber\\
y&=&\langle\hat{a}\hat{a}\rangle_0=
-\frac{A_0}{2\varepsilon}\coth\frac{\varepsilon}{2T},
\label{xy}
\end{eqnarray}
where $\varepsilon=\sqrt{B_0^2-|A_0|^2}$. We notice that at zero temperature
this method is nothing else but the approximation of the coherent squeezed
state (see, e.g., Ref.~\cite{mandel}) for the condensate mode.

It is more convenient to deal with the variables $x$ and $y$ rather than $A_0$
and $B_0$, since $\langle\hat{H}\rangle_0$ is easily expressed via $x$ and $y$
with Wick's theorem and $\d (\Omega_0-\langle\hat{H}_0\rangle_0)=-A_0\d y
-B_0\d x$. Hence, we come to the minimum conditions
\begin{equation}\label{condmin}
{\partial\langle\hat{H}\rangle_0}/{\partial y}=A_0, \quad
{\partial\langle\hat{H}\rangle_0}/{\partial x}=B_0,
\end{equation}
which should be solved in conjunction with
\begin{equation}\label{condcond}
\partial\langle\hat{H}\rangle_0/\partial z=0.
\end{equation}
For the Hartree Hamiltonian we have $\langle\hat{H}_{\rm h}\rangle_0
=\dst{\frac{V_0}{2V}}(2x^2+y^2+4xz+2zy+z^2)-\mu(z+x)$,
and Eqs.~(\ref{condmin}) and (\ref{condcond}) yield
\begin{eqnarray}
x&=& \frac{1-y/z}{4\sqrt{-y/z}}\coth\frac{\mu\sqrt{-y/z}}{T}-\frac{1}{2},
\nonumber\\
y&=&-\frac{1+y/z}{4\sqrt{-y/z}}\coth\frac{\mu\sqrt{-y/z}}{T},
\label{xyfinal}
\end{eqnarray}
where the asymptotic formula $V_0z/V\simeq\mu$ is utilized. The limits
$x/z\to0$ and $y/z\to0$ at $V\to\infty$ follow from Eq.~(\ref{asymp}), since
$z/V\simeq n_0$. If these limits were not fulfilled, the separation of the
condensate operator into quantum and $c$-number parts would have made no sense.
At zero temperature Eqs.~(\ref{xyfinal}) lead to $x\simeq -y\simeq
z^{1/3}/2^{4/3}$ and $x+1/2+y\simeq 2^{-5/3}z^{-1/3}$, and Eq.~(\ref{flmf}) yields
\begin{equation}\label{flh0}
\frac{\langle\delta \hat{N}_{0}^{2}\rangle}{N_{0}}
\simeq\frac{3}{2^{5/3}}\frac{1}{N_0^{1/3}}.
\end{equation}
This result was obtained for the Hartree model by another method in
Refs.~\cite{walls,dunn}. We note that the asymptotics~(\ref{flh0}) is in
agreement with Eq.~(\ref{flfinal}) at $T=0$, which reproduces only the exact
limit $\langle\delta\hat{N}_{0}^{2}\rangle/N_{0}=0$ for $N_0\to\infty$. At
nonzero temperature we obtain in the same manner $x\simeq -y\simeq
\sqrt{T/\mu}\sqrt{z}/2$ and $x+1/2+y\simeq T/(2\mu)$, and, hence,
\begin{equation}\label{flht}
\frac{\langle\delta\hat{N}_{0}^{2}\rangle}{N_{0}}
\simeq\frac{3}{2}\frac{T}{\mu}.
\end{equation}
The replacement $V_0\to4\pi\hslash^{2}a/m$, discussed in Sec.~\ref{sec:diff},
results in Eq.~(\ref{flfinal}) but with the factor $3/2$. Certainly, the
previous relation~(\ref{flfinal}) is valid, as based on the more general
considerations than the mean-field expression~(\ref{flht}). Note that the
method of the previous section, applied to the Hartree
Hamiltonian~(\ref{hartree}), also leads to the same correct
result~(\ref{flfinal}).

The mean-field scheme works even worse in the case of the
Hamiltonian~(\ref{heff}), for the subtle balance of the terms is broken in
Eqs.~(\ref{flmf}), (\ref{condmin}), and (\ref{condcond}). As a result, the term
$V_0=U_0$ vanishes in the limit $V\to\infty$, and the main contribution comes
in the condensate fluctuations from the term $U_1$. Note that the
approximations used in Ref.~\cite{glass} lead to the same effect. Thus, the
mean-field scheme, applied to the model Hamiltonian~(\ref{heff}), is not
consistent for calculating the fluctuations, because it does not reproduce the
correct answer~(\ref{flfinal}). Nevertheless, for a qualitative estimation of
the condensate fluctuations, it is quite right to make use of the mean-field
scheme, which reproduces the correct answer up to a factor of $3/2$. Note that
the Hartree Hamiltonian can be easily written down for the nonhomogeneous
system (see Sec.~2.8 in Ref.~\cite{walls}); the coefficients depend on the
condensate wave function (the eigenfunction of the one-body density matrix),
which is the solution of the Gross-Pitaevskii equation. So it is possible to
estimate qualitatively the condensate fluctuations for the trapped system by
means of of Eqs.~(\ref{flh0}) and (\ref{flht}).

\subsection{Next-to-leading order corrections for the fluctuations}
\label{sec:corr}

In this paper we study the condensate fluctuations within the single-mode
Hamiltonian~(\ref{heff}). As is stressed in Sec.~\ref{sec:model}, it means that
we neglect the correlations between the numbers of the particles in the
condensate and beyond it; that is, we put $[\langle\hat{N}_0\hat{N}_{\rm
out}\rangle -\langle\hat{N}_0\rangle \langle\hat{N}_{\rm out}\rangle]/V \simeq
0$ in the thermodynamic limit, where by definition $\hat{N}_{\rm
out}=\hat{N}-\hat{N}_{0} =\sum_{p\not=0} \hat{a}^{\dag}_{\bf p}\hat{a}_{\bf p}$
is the operator of particles beyond the Bose-Einstein condensate. In
Ref.~\cite{pit} an opposite idea was accepted that
$\langle\delta\hat{N}_0^2\rangle =\langle\delta\hat{N}_{\rm out}^2\rangle$ due
to the restriction $N={\rm const}$ in the canonical ensemble, but at the same
time, Bogoliubov's scheme, which does not conserve the number of particles, was
used there. This approach leads to nonthermodynamical fluctuations with
$\gamma=4/3$ in Eq.~(\ref{gamma}). As discussed in Sec.~\ref{sec:gen}, the
replacement of the condensate operators by the $c$-numbers implies that
Bogoliubov's terms are involved in the Hamiltonian. Such a procedure leads to
$[\hat{H},\hat{N}]\not=0$ and can change the accepted relation
$\langle\delta\hat{N}_0^2\rangle =\langle\delta\hat{N}_{\rm out}^2\rangle$.  On
the other hand, within the {\it conserving} scheme (when $[\hat{H},\hat{N}]=0$)
the Bogoliubov transformation  $\hat{b}_{\bf p}=u_{p}\hat{\alpha}_{\bf
p}+v_{p}\hat{\alpha}^{\dag}_{\bf -p}$ relates the creation
$\hat{\alpha}^{\dag}_{\bf p}$ and destruction $\hat{\alpha}_{\bf p}$
quasiparticle operators with not the particle operator $\hat{a}_{\bf p}$ but
with $\hat{b}_{\bf p}=\hat{a}_{\bf p}\hat{a}^{\dag}_{0}/\sqrt{N_{0}}$
\cite{bogzub,gard,cast}. Hence, the variance  of the operator
$\sum_{p\not=0}\hat{b}^{\dag}_{\bf p}\hat{b}_{\bf p}$ is no longer equal to the
variance of the number of noncondensate bosons $\langle\delta\hat{N}_{\rm
out}^2\rangle$. For this reason, the approach~\cite{pit} is implicitly based on
the assumption that $[\langle\hat{N}_{0}^2\hat{N}_{\rm out}^2\rangle
-\langle\hat{N}_{0}\hat{N}_{\rm out}\rangle^{2}]/N_{0}^{2}
\simeq\langle\hat{N}_{\rm out}^2\rangle-\langle\hat{N}_{\rm out}\rangle^{2}$ in
leading order.  A justification of this assumption is needed, since it may
occur that the main contribution to the l.h.s. comes from the term
$\langle\hat{N}_{\rm out}^2(\hat{N}_{0}^2-{N}_{0}^2)\rangle/N_{0}^{2}$, which
may be proportional to $V^{4/3}$. The question remains open what approximation
is more correct, the approximation of the quasi-independent mode or the
nonconserving approximation in conjunction with $\langle\delta
\hat{N}_0^2\rangle =\langle\delta\hat{N}_{\rm out}^2\rangle$. Note that our
result~(\ref{flfinal}) does not contradict to that of the
papers~\cite{idz,illu}, in which the fluctuations were investigated in the
canonical ensemble within the number-conserving simple scheme. We stress that
the Bogoliubov model, based on the $c$-number replacement, is consistent with
any value of $\gamma<2$, so all the approaches~\cite{pit} and \cite{idz,illu}
and ours do not contradict to the Bogoliubov's $c$-number replacement. Thus, we
are able to use the singe-mode Hamiltonian~(\ref{heff}) until a decisive answer
has been given what is the value of the correlator
$[\langle\hat{N}_{0}\hat{N}_{\rm out}\rangle-\langle\hat{N}_0\rangle
\langle\hat{N}_{\rm out}\rangle]/V$ within the number-conserving
scheme~\cite{note4}.

Let us formulate the hypothesis for the interacting Bose gas that in the
framework of {\it the number-conserving scheme} the relation
$\langle\hat{N}_0\hat{N}_{\rm out}\rangle/V\simeq\langle\hat{N}_0\hat{N}_{\rm
out}\rangle_c/V$ should be fulfilled; here, $\langle\cdots\rangle$ and
$\langle\cdots\rangle_c$ stand for the averages in the grand canonical and
canonical ensembles, respectively. This hypothesis is reasonable, since
transitions of bosons from the condensate state to the noncondensate ones and
back occur in the whole volume, and thus the boundary conditions seem to be of
no importance here. In addition, we have the relations
$\langle\hat{N}_{0}\hat{N}\rangle_c =N_{0}N$ and
$\langle\hat{N}_{0}\hat{N}\rangle -N_{0}N=T(\partial N_{0}/\partial\mu)_T$; the
latter can be derived by differentiating $\langle\hat{N}_{0}\rangle$ with
respect to the chemical potential (within the number-conserving scheme we do
not face the difficulties discussed in Sec.~\ref{sec:diff}). Thus, from the
accepted hypothesis we obtain
\begin{equation}\label{hip}
\frac{\langle\delta\hat{N}^2_{0}\rangle}{N_{0}}=\frac{T}{n_{0}}\bigg(\frac{\partial n_{0}}{\partial\mu}\bigg)_T
+\frac{\langle\delta\hat{N}^2_{0}\rangle_c}{N_{0}}.
\end{equation}
Here the derivative $(\partial\mu/\partial n_0)_T$ is not related to the
formula~(\ref{dmudn0}), which concerns the single-mode
Hamiltonian~(\ref{heff}), because the averages in Eq.~(\ref{hip}) correspond to
the full Hamiltonian with the pairwise potential; one should calculate it from
the thermodynamic expression for the chemical potential. In particular, in the
temperature region $n_0U^{(0)}(0)\ll T\ll T_{\text{c}}$ [here $T_{\text{c}}$ is
the transition temperature of the ideal Bose gas and
$U^{(0)}(0)=4\pi\hslash^{2}a/m$] we have
\[
n_0\bigg(\frac{\partial\mu}{\partial n_0}\bigg)_T \simeq n_0U^{(0)}(0)
\bigg[1-6\sqrt{\pi}\frac{(n_0a^3)^{1/2}T}{n_0U^{(0)}(0)}\bigg],
\]
where $(n_0a^3)^{1/2}T/ \big[n_0U^{(0)}(0)\big]\ll 1$ (see, e.g.,
Ref.~\cite{shi}). On the other hand, the number-conserving approach of
Ref.~\cite{idz} yields
\[
\frac{\langle\delta\hat{N}_{0}^{2}\rangle_c}{N_0}\simeq
2\sqrt{2}\sqrt{\pi}(n_0a^3)^{1/2} \bigg(\frac{T}{n_0U^{(0)}(0)}\bigg)^2
\vspace*{2mm}
\]
in that temperature region~\cite{note5}. Thus, with the proposed
equation (\ref{hip}) we obtain the relation
\begin{eqnarray}
\frac{\langle\delta\hat{N}^2_{0}\rangle}{N_{0}}\! \simeq\!\frac{T}{n_{0}U^{(0)}(0)}
\bigg[1+(6+2\sqrt{2})\sqrt{\pi}\frac{(n_0a^3)^{1/2}T}{n_0U^{(0)}(0)}\bigg],
\label{correction}
\end{eqnarray}
which contains the correction term in comparison with Eq.~(\ref{flfinal}).

\section{Summary}
\label{sec:dis}

Starting from the approximation of the quasi-independent mode, we derive the single-mode
Hamiltonian~(\ref{heff}) and obtain the condensate fluctuations~(\ref{flfinal}) for the grand canonical ensemble in the
dilute limit. This relation is derived beyond the mean-field approximation. The mean-field scheme, applied to the
Hamiltonian~(\ref{heff}), leads to incorrect results. For the Hartree Hamiltonian~(\ref{hartree}), the mean-field
approximation results in Eq.~(\ref{flht}) at nonzero temperature, which differs from the correct relation~(\ref{flfinal})
by a factor of $3/2$. With the help of the proposed hypothesis and the estimations of Ref.~\cite{idz}, the next-to-leading
term is obtained for the condensate fluctuations~(\ref{correction}).

This work was supported by RFBR grant No. 01-02-17650.



\begin{thebibliography}{99}
\bibitem{note3} These relations are obtained at zero temperature, but the temperature corrections
are not important for estimating the leading term in Eq.~(\ref{disp}).
\bibitem{classif} We employ the following classification of the short-range
pairwise potentials. The potential is of the weak-coupling type if it is
integrable and proportional to a small parameter~(the coupling constant). In
this case the Born approximation works well when applied to the ordinary
two-body problem. Otherwise the potential is called singular. Finally, we deal
with the strongly singular potential (the strong-coupling regime) if the
individual terms in the Born series for the scattering amplitude do not exist
due to a strong repulsion at short particle separations. For example, 
Lennard-Jones and hard-sphere potentials belong to that type.
\bibitem{note} We notice that the divergence may arise due to the
approximations made when deriving Eq.~(\ref{disp}), so it is possible that it
is not inherent in the model Hamiltonian~(\ref{heff}) itself.
\bibitem{note5} Note that, in spite of the simplicity of the model~\cite{idz},
it yields a qualitatively valid expression for  condensate depletion in the
mentioned temperature region. So the expansion of Eq.~(7) in Ref.~\cite{idz}
reproduces Eq.~(3.53) of Ref.~\cite{shi} up to a factor of $\sqrt{2}$.
\end{thebibliography}
\end{document}